# A Set of Formulae on Fractal Dimension Relations and Its Application to Urban Form

Yanguang Chen

(Department of Geography, College of Urban and Environmental Sciences, Peking University, Beijing 100871, PRC. E-mail: chenyg@pku.edu.cn)

**Abstract**: The area-perimeter scaling can be employed to evaluate the fractal dimension of urban boundaries. However, the formula in common use seems to be not correct. By means of mathematical method, a new formula of calculating the boundary dimension of cities is derived from the idea of box-counting measurement and the principle of dimensional consistency in this paper. Thus, several practical results are obtained as follows. First, I derive the hyperbolic relation between the boundary dimension and form dimension of cities. Using the relation, we can estimate the form dimension through the boundary dimension and *vice versa*. Second, I derive the proper scales of fractal dimension: the form dimension comes between 1.5 and 2, and the boundary dimension comes between 1 and 1.5. Third, I derive three form dimension values with special geometric meanings. The first is 4/3, the second is 3/2, and the third is $1+2^{1/2}/2 \approx 1.7071$. The fractal dimension relation formulae are applied to China's cities and the cities of the United Kingdom, and the computations are consistent with the theoretical expectation. The formulae are useful in the fractal dimension estimation of urban form, and the findings about the fractal parameters are revealing for future city planning and the spatial optimization of cities.

**Key words**: Allometric scaling; Area-perimeter scaling; Fractal; Structural fractal; Textural fractal; Form dimension; Boundary dimension; Urban form; Urban shape

# 1 Introduction

An urban landscape in a digital map is a kind of irregular pattern consisting of countless fragments. This landscape reminds us of fractals, which comprise form, chance, and dimension (Mandelbrot, 1977; Mandelbrot, 1982). Empirically, urban form can be characterized with fractal geometry. A number of studies showed that fractal theory is a powerful tool for urban spatial analysis (see e.g. Ariza-Villaverde *et al*, 2013; Batty and Longley, 1994; Benguigui *et al*, 2000; Chen and Feng, 2012; Frankhauser, 1998; De Keersmaecker *et al*, 2003; Lu and Tang, 2004; Murcio and Rodríguez-Romo, 2009; Terzi and Kaya, 2011; Thomas *et al*, 2007; Thomas *et al*, 2008; White and Engelen, 1994). In urban studies, fractal dimension is a basic and useful measure of urban shape and structure and it is employed to describe urban growth and form (see. e.g. Batty and Longley, 1987; Batty and Longley, 1988; Benguigui *et al*, 2006; Feng and Chen, 2010; Frankhauser, 1994; Murcio and Rodríguez-Romo, 2011; Shen, 2002; Thomas *et al*, 2010; White and Engelen, 1993). Recent twenty years, many of our theories in urban geography have been reinterpreted using ideas from fractals. However, despite various works on city fractals, we have little research on the fractal dimension of cities itself, the regularity of fractal dimension change, and relations between different fractal dimensions (Chen, 2012; Chen, 2013).

One of efficient approach to revealing the regularity and relation of different fractal dimensions of cities is to combine different methods of defining fractal dimension. By doing so, we can find the inherent relationships between different fractal parameters. Using these relationships, we can derive some useful formulae of fractal dimension estimation. In the practice of determining fractal dimension, we can often reach the same goal by different routes. For example, for the fractal dimension of urban boundary, we can estimate it with not only the area-perimeter scaling, but also the perimeter-scale relation (Batty and Longley, 1994; Chen, 2011; Longley and Batty, 1989a; Longley and Batty, 1989b; Wang *et al*, 2005). Thus we can construct a system of scaling equations. Fractal dimension relations can be found by finding the solution to the equation set. By means of the dimension relations, we can bring to light the connections between different fractal dimensions and the regularity of fractal dimension change.

This paper is devoted to revealing the mathematical and numerical relationships between the fractal dimension of urban structure (pattern) and that of urban texture (boundary), and showing the numerical regularity of fractal dimension change based on the box-counting method. The rest parts are organized as follows. First, the hyperbolic relation between the fractal dimension of urban form and that of urban boundary are derived, and a set of fractal dimension formulae are presented. A wrong definition of fractal dimension based on the area-perimeter relation that is employed by usage to estimate the fractal dimension is reclaimed from the traditional point of view (Section 2). Second, as examples, the fractal dimension relation equations are applied to Chinese cities and British cities to show the effect of the new formulae (Section 3). Third, several questions are discussed and the regularity of the box-counting dimensions of cities is partially uncovered (Section 4). Finally, the paper is concluded by summarizing the mains of this study.

## 2 Fractal dimension equations

### 2.1 Geometric measure relation and fractal dimension

Fractals and fractal dimension can be understood from the viewpoint of geometric measure relation. According to the axiom of dimensional consistency, a geometric measure $X$ is proportional to another measure $Y$ if and only if the dimension values of the two measures are equal to one another (Chen, 2013; Lee, 1989; Mandelbrot, 1982; Takayasu, 1990). If the two dimensions is inconsistent, we cannot obtain a proportional relation such as $Y \propto X$ or $Y=kX$, where $k$ denotes a proportionality constant, and "$\propto$" means "be in proportion to". If we want to construct proportional relations between different measures, we must make the dimensions consistent. Therefore, for a length, $L$, an area, $A$, a volume, $V$, and a general spatial quantity (any "mass"), $M$, we have a measure relation such as

$$L^{1/1} \propto A^{1/2} \propto V^{1/3} \propto M^{1/d_f}, \tag{1}$$

where $d_f$ refers to a general dimension (Mandelbrot, 1982; Takayasu, 1990). For a Euclidean object, we have $d_f=0$ for a point, $d_f=1$ for a line, $d_f=2$ for a plane, and $d_f=3$ for a cube. However, for a fractal object, the $d_f$ value will vary from 0 to 3. Based on equation (1), an area-perimeter scaling relation was given as follows

$$A \propto P^{2/D_l} = P^{\sigma}, \tag{2}$$

where $A$ refers to area (say, urban area), and $P$ to perimeter (say urban boundary length). Equation (2) is in fact an allometric scaling relation (Chen, 2010). What is called "allometric scaling relation" implies that the ratio of the change rate of one measure to that of another measure in a system is a constant. Actually, the allometric scaling is a generalized allometric growth (Batty and Longley, 1994; Chen, 2010; Lee, 1989). In other words, equation (2) is valid only for the allometric growth. For a growing fractal object, the power law should be substituted with an inverse power law. Obviously, we have an allometric scaling exponent

$$\sigma = \frac{d}{D_l} = \frac{2}{D_l},\qquad(3)$$

where $d=2$ denotes a Euclidean dimension, and $D_l$ used to be regarded as the fractal dimension of a closed irregular curve bounding a plane area (see e.g. Feder, 1988; Mandelbrot, 1982; Lung and Mu, 1988). For cities, $D_l$ was considered to be the fractal dimension of urban boundary (Batty and Longley, 1994; Chen, 2010; Longley *et al*, 1991; Wang *et al*, 2005). However, $D_l$ is not really a boundary dimension, but a ratio of two fractal dimensions (Benguigui *et al*, 2006; Chen, 2010; Cheng, 1995; Imre, 2006; Imre and Bogaert, 2004). In practice, the fractal parameter $D_l$ can be termed *quasi-dimension* of a fractal line. What is more, equation (1) is not completely correct for a fractal object (e.g. Koch island), and it should be amended in terms of fractal notion.

Now, a question arises and remains to be answered. If $D_l$ is not the real boundary dimension, how can we find the fractal dimension through the area-perimeter scaling? This seems to be a problem that defies solution. In fact, the problem can be solved by means of scaling equations. The key of this study process is the box-counting method of fractal dimension estimation. Through box counting, the fractal dimension can not only be directly calculated by the area/perimeter-scale relations (see equations (5) and (6)), but also be indirectly estimated by using the area-perimeter scaling (see equation (9)). Thus we have a system of scaling equations. By resolving the equations set, we can derive the relation between the different fractal parameters; from the relation it follows a series of useful formulae on fractal dimension relations.

## 2.2 Derivations of fractal dimension relations

Suppose there exists a city with fractal form, and the fractal dimension is examined in a 2-dimensional Euclidean space. The fractal form indicates the self-similar shape and structure of a city. Two fractal dimensions can be employed to characterize the urban form. One is form dimension, which is defined with the area-scale scaling based on a 2-dimensional fractal initiator, and the other is the boundary dimension, which is defined with the perimeter-scale scaling based on a 1-dimensional fractal initiator. To evaluate the fractal dimensions of urban form, we can make an evenly-spaced grid to lay on it by means of a digital map. Each cell represents a general "box" in the grid system. Count how many "boxes" are required to cover an urban area or boundary. Changing the side length of boxes, $\varepsilon$, leads to change of nonempty boxes number, $N(\varepsilon)$. Since the urban form is self-similar, the measure relation between the linear scale of the boxes and the number of the least nonempty boxes will follow an inverse power law such as

$$N(\varepsilon) = N_1 \varepsilon^{-D}, \qquad (4)$$

where $N_1$ refers to the proportionality coefficient, and $D$, to the fractal dimension of urban form.

The box-counting method can be utilized to estimate the length of urban boundary and the urban area within the close urban boundary (Benguigui *et al*, 2000; Wang *et al*, 2005). A complete urban boundary is termed *urban envelope* (Batty and Longley, 1994; Longley *et al*, 1991). In fact, we can use the functional box-counting method (Lovejoy *et al*, 1987). Adopting a grid system to cover an urban image on a digital map, we will have a pattern of regularly spaced horizontal and vertical lines forming squares or rectangles on the map (Feng and Chen, 2010; Shen, 2002). Two kinds of numbers can be given by counting the nonempty "box"—the cells/squares with parts of urban figures inside. The total number of the "boxes" including the urban boundary is notated as $N_b(\varepsilon)$, and the totality of the "boxes" including the urban patches within the urban envelope notated as $N_f(\varepsilon)$. Consequently, the urban perimeter can be estimated by $P(\varepsilon)=N_b(\varepsilon)\varepsilon$, while the urban area can be estimated by $A(\varepsilon)=N_f(\varepsilon)\varepsilon^2$. Changing the linear size of the "boxes", $\varepsilon$, yields different values of $A(\varepsilon)$ and $P(\varepsilon)$, which are larger the shorter the linear scale $\varepsilon$ is. In theory, if $\varepsilon$ becomes small enough, $A(\varepsilon)$ and $P(\varepsilon)$ will represent the urban area and perimeter, respectively.

The process of deriving the fractal dimension relations are as follows. Appling equation (4) to an urban figure within its urban envelope on a digital map yields

$$N_f(\varepsilon) = N_{f1} \varepsilon^{-D_f}, \qquad (5)$$

where $N_{f1}$ is the proportionality coefficient, and $D_f$, to the fractal dimension of urban form, termed *form dimension*. Form dimension is a kind of *structural fractal dimension* (Addison, 1997; Chen and Zhou, 2006; Kaye, 1989). Further, apply equation (2) to an urban boundary, without taking urban area inside the perimeter into account, yields

$$N_b(\varepsilon) = N_{b1}\varepsilon^{-D_b}, \tag{6}$$

where $N_{b1}$ is the proportionality constant, and $D_b$, to the fractal dimension of urban boundary, termed *boundary dimension*. Boundary dimension is a type of *textural fractal dimension* (Addison, 1997; Chen and Zhou, 2006; Kaye, 1989). The structural dimension is used to describe urban form comprising points, lines and patches, while the textural dimension is used to describe urban boundary or the interurban/intraurban fractal curves comprising line segments.

Because of fractal property of cities, urban area and urban perimeter are not fixed. Apparently, the urban area can be estimated as

$$A(\varepsilon) = N_f(\varepsilon)\varepsilon^2 = A_1\varepsilon^{2-D_f}, \tag{7}$$

where $A_1$ is a proportionality coefficient. Equation (7) is a power law indicative of direct proportions because $D_f<2$. The smaller the linear scale $\varepsilon$ is, the closer the $A(\varepsilon)$ value is to the real urban area. The urban perimeter can be given by

$$P(\varepsilon) = N_b(\varepsilon)\varepsilon = P_1\varepsilon^{1-D_b}, \tag{8}$$

where $P_1$ is also a proportionality coefficient. Equation (8) is a power law indicating inverse proportions because $D_b>1$. The smaller the linear scale $\varepsilon$ is, the closer the $P(\varepsilon)$ value is to the real circumference. From equations (7) and (8) it follows

$$A(\varepsilon) = A_1 P_1^{(2-D_f)/(D_b-1)} P(\varepsilon)^{(2-D_f)/(1-D_b)} \propto P(\varepsilon)^{-\sigma}, \tag{9}$$

where $\sigma>0$ denotes a scaling exponent. Thus we have

$$\sigma = \frac{2-D_f}{D_b-1}. \tag{10}$$

This suggests that the scaling exponent of urban area and perimeter depends on the form dimension and boundary dimension of a city. For a Euclidean geometrical object, equations (9) and (10) will be invalid.

The fractal measure relation between urban area and perimeter can be derived from the principle of dimension consistency. According to equation (1), we have an area-perimeter scaling such as

$$A(\varepsilon) \propto P(\varepsilon)^{-D_f/D_b} \propto P(\varepsilon)^{-\sigma}, \tag{11}$$

in which

$$\sigma = \frac{D_f}{D_b}. \quad (12)$$

Note that the function of positive power is replaced by the function of negative power. Combining equation (10) and equation (12) yields

$$\sigma = \frac{2 - D_f}{D_b - 1} = \frac{D_f}{D_b}, \quad (13)$$

which gives a hyperbolic relation between the form dimension and the boundary dimension as below:

$$\frac{1}{D_b} = 2 - \frac{2}{D_f}. \quad (14)$$

This suggests, given $D_f$, it follows that

$$D_b = \frac{D_f}{2(D_f - 1)}, \quad (15)$$

which indicates that $D_f \neq 1$. On the other hand, given $D_b$, it follows that

$$D_f = \frac{2D_b}{2D_b - 1}, \quad (16)$$

which indicates that $D_b \neq 1/2$. Further, in terms of equations (3), (12), and (14), given $D_l$ or $\sigma$, it follows that

$$D_b = \frac{2 + \sigma}{2\sigma} = \frac{1 + D_l}{2}, \quad (17)$$

and

$$D_f = \sigma D_b = \frac{2 + \sigma}{2} = 1 + \frac{1}{D_l}, \quad (18)$$

which suggest that there is a linear relation between $D_l$ and $D_b$, and a hyperbolic relation between $D_l$ and $D_f$.

To sum up, we can estimate the form dimension and boundary dimension by using the scaling exponent $\sigma$ or the quasi-dimension $D_l$. Equation (14) gives the theoretical relation between the form dimension $D_f$ and boundary dimension $D_b$, and equations (17) and (18) give a pair of practical formulae of fractal dimension estimation for urban boundary and form. In empirical studies, it is difficult to estimate the fractal dimensions $D_b$ and $D_f$, but easy to evaluate the allometric scaling exponents $D_l$ or $\sigma$ (Batty and Longley, 1988; Chen, 2010). Equations (17) and

(18) are very useful for us to estimate $D_b$ and $D_f$ indirectly. In next section, based on the area-perimeter scaling and the box-counting method, the formulae will be applied to the cities in the real world to show how to estimate the form and boundary dimensions.

## 3 Application to urban form of real cities

### 3.1 Data and method

The area–perimeter relation is a widely used method to estimate the perimeters' fractal dimension of self-similar shapes which are embedded into a 2-dimensional Euclidean space. In previous literature, we can find the quasi-dimension $D_l$ or the scaling exponent $\sigma$ or the reciprocal of the scaling exponent $1/\sigma$, but we barely find the form dimension $D_f$ and boundary dimension $D_b$ from the area-perimeter scaling based on the box-counting method. Actually, using the fractal dimension formulae, we can convert the quasi-dimension $D_l$ or the scaling exponent $\sigma$ into the form dimension $D_f$ and boundary dimension $D_b$. As an example, the method is applied to China's cities now. Wang *et al* (2005) estimated the values of the quasi-fractal dimension of urban boundary of 31 China's megacities in 1990 and 2000. The original datasets came from the database of the national resources and environment of the Institute of Geographic Sciences and Natural Resources Research (IGSNRR), Chinese Academy of Science (CAS), China. The database was founded by means of the technologies of remote sensing (RS) and geographical information systems (GIS) (Chen, 2011).

A city differs to some extent from a real fractal, and it can be treated as a kind of prefractal (Addison, 1997). For a mathematical fractal, the area-perimeter scaling follows an inverse power law, equation (9). However, for a city, the fractal measure relation is replaced with an allometric scaling relation, and thus the area-perimeter scaling follows a power law, equation (2). Suppose that the relationship between urban area and perimeter follows the allometric scaling law (Chen, 2010). Taking the logarithm on both sides of equation (2) gives

$$\ln P(\varepsilon) = C + \frac{D_l}{2} \ln A(\varepsilon) = C + \frac{1}{\sigma} \ln A(\varepsilon), \tag{19}$$

where $C$ refers to a constant. A scaling exponent can be estimated with the least squares method. Based on equation (19), a hybrid approach combining the area-perimeter scaling and box-counting method was utilized to estimate the quasi-dimensions of China's cities by Wang *et al* (2005). This method shares the similar principles with the cell-count method developed by Longley and Batty (1989a; 1989b). In fact, Wang et al (2005) adopted 19 sets of grids with different linear scales to cover the built-up area of the 31 cities. In other words, for each city, the linear scale of squares changes 19 times. By the resolution of digital maps, the lower limit of squares for the grid systems corresponds to regional units of 200 (m)×200 (m) on the surface of the earth. With the aid of GIS and advanced programming language, Wang et al (2005) estimated the scaling exponents of the 31 mage-cities, i.e., the $\sigma$ values, in 1990 and 2000. Thus, the quasi-dimension of urban boundary can be evaluated with the formula $D_l=2/\sigma$ (Table 1).

## 3.2 Results

It is easy to compute the boundary dimension and form dimension of a city using the formulae presented in Subsection 2.2. The scaling exponent $\sigma$ and the quasi-dimension $D_l$ can be determined with the traditional approach, which gives nothing about the boundary dimension $D_b$ and the form dimension $D_f$. Using equation (17), we can estimate the boundary dimensions of the 31 cities for 1990 and 2000; using equation (18), we can estimate the form dimensions of these cities for the two years (Table 1).

As indicated above, the quasi-dimension $D_l$ used to be mistaken as the boundary dimension $D_b$. In essence, the two fractal parameters are different from one another. In Table 1, for 1990, the maximum of $D_l$ is 1.748, the minimum is 1.300, and the average is about 1.483, which is close to 1.5; for 2000, the maximum of $D_l$ is 1.742, the minimum is 1.278, and the mean is around 1.454. The values are too high where the boundary dimension is concerned. A fractal boundary bears an analogy to the Koch curve with a dimension about 1.262. However, in light of the revised results formulated with equations (17) and (18), the boundary dimension ranges from 1.150 to 1.374 in 1990, and vary from 1.139 to 1.371 in 2000; Accordingly, the form dimension ranges from 1.572 to 1.742 in 1990, and vary from 1.574 to 1.782 in 2000. The values of $D_b$ come between 1 and 1.5, and its mean decrease from 1.242 in 1990 to 1.227 in 2000. The $D_f$ values fall into 1.5 and 2, and the mean increase from 1.677 in 1990 to 1.691 in 2000. If $D_l$ were the boundary dimension, the result would be strange. In contrast, the corrective values are reasonable and acceptable.

**Table 1 The scaling exponent of China's 31 megacities in 1990 and 2000 and the corresponding**

**results of fractal dimension estimation**

| City/Statistics | 1990 | | | | 2000 | | | |
|---|---|---|---|---|---|---|---|---|
| | $D_l$ | $\sigma$ | $D_b$ | $D_f$ | $D_l$ | $\sigma$ | $D_b$ | $D_f$ |
| Anshan | 1.469 | 1.361 | 1.235 | 1.681 | 1.380 | 1.449 | 1.190 | 1.725 |
| Beijing | 1.502 | 1.332 | 1.251 | 1.666 | 1.444 | 1.385 | 1.222 | 1.693 |
| Changchun | 1.404 | 1.425 | 1.202 | 1.712 | 1.401 | 1.428 | 1.201 | 1.714 |
| Changsha | 1.532 | 1.305 | 1.266 | 1.653 | 1.526 | 1.311 | 1.263 | 1.655 |
| Chengdu | 1.676 | 1.193 | 1.338 | 1.597 | 1.674 | 1.195 | 1.337 | 1.597 |
| Chongqing | 1.505 | 1.329 | 1.253 | 1.664 | 1.446 | 1.383 | 1.223 | 1.692 |
| Dalian | 1.489 | 1.343 | 1.245 | 1.672 | 1.474 | 1.357 | 1.237 | 1.678 |
| Fushun | 1.411 | 1.417 | 1.206 | 1.709 | 1.366 | 1.464 | 1.183 | 1.732 |
| Guangzhou | 1.403 | 1.426 | 1.202 | 1.713 | 1.544 | 1.295 | 1.272 | 1.648 |
| Guiyang | 1.748 | 1.144 | 1.374 | 1.572 | 1.742 | 1.148 | 1.371 | 1.574 |
| Hangzhou | 1.599 | 1.251 | 1.300 | 1.625 | 1.565 | 1.278 | 1.283 | 1.639 |
| Harbin | 1.369 | 1.461 | 1.185 | 1.730 | 1.307 | 1.530 | 1.154 | 1.765 |
| Jilin | 1.424 | 1.404 | 1.212 | 1.702 | 1.432 | 1.397 | 1.216 | 1.698 |
| Jinan | 1.433 | 1.396 | 1.217 | 1.698 | 1.463 | 1.367 | 1.232 | 1.684 |
| Kunming | 1.588 | 1.259 | 1.294 | 1.630 | 1.472 | 1.359 | 1.236 | 1.679 |
| Lanzhou | 1.482 | 1.350 | 1.241 | 1.675 | 1.471 | 1.360 | 1.236 | 1.680 |
| Nanchang | 1.454 | 1.376 | 1.227 | 1.688 | 1.502 | 1.332 | 1.251 | 1.666 |
| Nanjing | 1.569 | 1.275 | 1.285 | 1.637 | 1.494 | 1.339 | 1.247 | 1.669 |
| Qingdao | 1.377 | 1.452 | 1.189 | 1.726 | 1.305 | 1.533 | 1.153 | 1.766 |
| Qiqihar | 1.355 | 1.476 | 1.178 | 1.738 | 1.340 | 1.493 | 1.170 | 1.746 |
| Shanghai | 1.481 | 1.350 | 1.241 | 1.675 | 1.422 | 1.406 | 1.211 | 1.703 |
| Shenyang | 1.300 | 1.538 | 1.150 | 1.769 | 1.278 | 1.565 | 1.139 | 1.782 |
| Shijiazhuang | 1.571 | 1.273 | 1.286 | 1.637 | 1.466 | 1.364 | 1.233 | 1.682 |
| Taiyuan | 1.554 | 1.287 | 1.277 | 1.644 | 1.538 | 1.300 | 1.269 | 1.650 |
| Tangshan | 1.500 | 1.333 | 1.250 | 1.667 | 1.456 | 1.374 | 1.228 | 1.687 |
| Tianjin | 1.376 | 1.453 | 1.188 | 1.727 | 1.356 | 1.475 | 1.178 | 1.737 |
| Urumchi | 1.447 | 1.382 | 1.224 | 1.691 | 1.441 | 1.388 | 1.221 | 1.694 |
| Wuhan | 1.475 | 1.356 | 1.238 | 1.678 | 1.494 | 1.339 | 1.247 | 1.669 |
| Xian | 1.461 | 1.369 | 1.231 | 1.684 | 1.366 | 1.464 | 1.183 | 1.732 |
| Zhengzhou | 1.506 | 1.328 | 1.253 | 1.664 | 1.426 | 1.403 | 1.213 | 1.701 |
| Zibo | 1.525 | 1.311 | 1.263 | 1.656 | 1.493 | 1.340 | 1.247 | 1.670 |
| Maximum value | 1.748 | 1.538 | 1.374 | 1.769 | 1.742 | 1.565 | 1.371 | 1.782 |
| Minimum value | 1.300 | 1.144 | 1.150 | 1.572 | 1.278 | 1.148 | 1.139 | 1.574 |
| Average | 1.483 | 1.353 | 1.242 | 1.677 | 1.454 | 1.381 | 1.227 | 1.691 |

**Note**: The values of scaling exponent $D_l$ come from Wang et al, 2005, and the form dimension $D_f$ and the boundary dimension $D_b$ are estimated with equations (17) and (18).

## 3.3 Other cases

The cities discussed above are all megacities in China. In fact, the fractal parameter equations can also be employed to estimate the fractal dimension of other type of cities, for example, the mining cities. Using the method developed by Wang et al (2005), Song et al (2012) calculated the quasi-dimension values ($D_l$) of 33 mining cities of China, including megacities, large cities, medium-sized cities, and small cities. Applying equations (17) and (18) to the results from Song et al (2012) yields the form dimension ($D_f$) and boundary dimension ($D_b$) of the 33 cities in 2006 (Table 2). The boundary dimension values come between 1.126 and 1.299, the average value is about 1.227; the form dimension values range from 1.626 to 1.799, the average is around 1.691. Where means are concerned, the results of the 33 mining cities in 2006 are very close to those of the 31 megacities in 2000 (Table 1, Table 2).

Table 2 The scaling exponent of China's 33 mining cities in 2006 and the corresponding results of fractal dimension estimation

| City/Statistic | $D_l$ | $R^2$ | $\sigma$ | $D_b$ | $D_f$ |
| --- | --- | --- | --- | --- | --- |
| Anshan | 1.441 | 0.978 | 1.388 | 1.221 | 1.694 |
| Baiyin | 1.283 | 0.985 | 1.559 | 1.142 | 1.779 |
| Benxi | 1.593 | 0.980 | 1.255 | 1.297 | 1.628 |
| Daqing | 1.555 | 0.978 | 1.286 | 1.278 | 1.643 |
| Datong | 1.413 | 0.983 | 1.415 | 1.207 | 1.708 |
| Dongying | 1.443 | 0.982 | 1.386 | 1.222 | 1.693 |
| Fushun | 1.526 | 0.975 | 1.311 | 1.263 | 1.655 |
| Fuxin | 1.448 | 0.983 | 1.381 | 1.224 | 1.691 |
| Hebi | 1.498 | 0.981 | 1.335 | 1.249 | 1.668 |
| Hegang | 1.510 | 0.981 | 1.325 | 1.255 | 1.662 |
| Huaibei | 1.487 | 0.979 | 1.345 | 1.244 | 1.672 |
| Huainan | 1.525 | 0.982 | 1.311 | 1.263 | 1.656 |
| Jixi | 1.574 | 0.979 | 1.271 | 1.287 | 1.635 |
| Jinchang | 1.251 | 0.989 | 1.599 | 1.126 | 1.799 |
| Jincheng | 1.309 | 0.986 | 1.528 | 1.155 | 1.764 |
| Karamay | 1.284 | 0.989 | 1.558 | 1.142 | 1.779 |
| Liaoyuan | 1.509 | 0.977 | 1.325 | 1.255 | 1.663 |
| Liupanshui | 1.444 | 0.975 | 1.385 | 1.222 | 1.693 |
| Maanshan | 1.337 | 0.987 | 1.496 | 1.169 | 1.748 |
| Panzhihua | 1.546 | 0.986 | 1.294 | 1.273 | 1.647 |
| Panjin | 1.509 | 0.981 | 1.325 | 1.255 | 1.663 |
| Pingdingshan | 1.517 | 0.980 | 1.318 | 1.259 | 1.659 |
| Pingxiang | 1.358 | 0.982 | 1.473 | 1.179 | 1.736 |
| Puyang | 1.476 | 0.981 | 1.355 | 1.238 | 1.678 |
| Qitaihe | 1.546 | 0.988 | 1.294 | 1.273 | 1.647 |
| Shizuishan | 1.256 | 0.989 | 1.592 | 1.128 | 1.796 |
| Shuangyashan | 1.597 | 0.984 | 1.252 | 1.299 | 1.626 |
| Shuozhou | 1.326 | 0.981 | 1.508 | 1.163 | 1.754 |
| Songyuan | 1.468 | 0.981 | 1.362 | 1.234 | 1.681 |
| Tangshan | 1.476 | 0.982 | 1.355 | 1.238 | 1.678 |
| Tongchuan | 1.426 | 0.982 | 1.403 | 1.213 | 1.701 |
| Wuhai | 1.557 | 0.981 | 1.285 | 1.279 | 1.642 |
| Yangquan | 1.489 | 0.983 | 1.343 | 1.245 | 1.672 |
| Maximum value | 1.597 | 0.989 | 1.599 | 1.299 | 1.799 |
| Minimum value | 1.251 | 0.975 | 1.252 | 1.126 | 1.626 |
| Average | 1.454 | 0.982 | 1.382 | 1.227 | 1.691 |

**Note**: The values of scaling exponent $D_l$ come from Song et al, 2012, and the form dimension $D_f$ and the boundary dimension $D_b$ are estimated with equations (17) and (18).

The formulae of fractal dimension relation can also be utilized to estimate the fractal dimension of the cities in European countries such as the United Kingdom (UK). The first case is Cardiff, the capital and largest city of Wales, in the southeast part of the country on Bristol Channel. By means of equation (6) and digital maps, Longley and Batty (1989b) once estimated the boundary dimension of Cardiff in 1886, 1901, 1922, and 1949. However, we know nothing about the form dimension. Using equation (14), we can easily estimate the form dimension (Table 3). The second case is Swindon, the municipal borough of south-central England, which is to the east-northeast of Bristol. By means of a digital map and the area-perimeter scaling relation similar to equation (7), Batty and Longley (1988) once estimated the quasi-dimension of different urban land use. By using equations (13) and (14), we can estimate the boundary dimension and form dimension (Table 4).

Table 3 The boundary dimension and the corresponding form dimension of Cardiff in four years

| Year | $D_b$ | $R^2$ | $D_f$ |
|---|---|---|---|
| 1886 | 1.267 | 0.953 | 1.652 |
| 1901 | 1.200 | 0.967 | 1.714 |
| 1922 | 1.209 | 0.957 | 1.705 |
| 1949 | 1.274 | 0.985 | 1.646 |

Table 4 The quasi-dimension, the corresponding boundary dimension and form dimension of Swindon, 1981.

| Land use | $D_l$ | $D_b$ | $D_f$ |
|---|---|---|---|
| Residential land use | 1.3310 | 1.1655 | 1.7513 |
| Commercial-industrial land use | 1.4779 | 1.2390 | 1.6766 |
| Educational land use | 0.5694* | 0.7847 | 2.7562 |
| Transport land use | 1.4471 | 1.2236 | 1.6910 |
| Open space | 1.2435 | 1.1218 | 1.8042 |
| All land uses | 1.2961 | 1.1481 | 1.7715 |

*Note: Because the sample is too small, the fractal parameter estimation of the educational land use is not proper.

## 4 Questions and discussion

The quasi-fractal dimension of urban boundary $D_l$ used to be confused with the real boundary dimension $D_b$. The main result of this paper is deriving a set of formulae, which act as a new approach to estimating the fractal dimension of urban form and boundaries. In particular, the fractal parameter relations provide us with a new theoretical way of understanding fractal dimension of urban patterns. According to equation (14), the relation between the form dimension, $D_f$, and the boundary dimension, $D_b$, is hyperbolic (Figure 1). The higher form dimension suggests the lower boundary dimension, and *vice versa*. As examples, a series of numerical values of the two kinds of fractal dimension are listed in Table 5. Several inferences can be drawn as follows. First, the form dimension cannot equal 1, or else the boundary dimension will be infinity, which suggests a meaningless value of the boundary dimension. Second, the form dimension must be greater than 4/3≈1.333 ($D_f$>4/3), otherwise the boundary dimension will be greater than 2 in theory. Third, if the form dimension equals 2 ($D_f$=2), the boundary dimension will equal 1 ($D_b$=1). This suggests that the form dimension cannot be equal to 2, and thus the above formulae are invalid for Euclidean geometry. Fourth, the fractal dimension $D_f$=1.5 is a critical value. If the form dimension equals 1.5 ($D_f$=3/2), we will have the boundary dimension equal to 1.5 ($D_b$=3/2) and *vice versa*.

Table 5 The numerical relation between the form dimension and boundary dimension

| $D_f$ | $D_b$ | $D_f$ | $D_b$ | $D_f$ | $D_b$ | $D_f$ | $D_b$ |
|---|---|---|---|---|---|---|---|
| 1.000 | ∞ | **1.300** | **2.167** | 1.550 | 1.409 | 1.800 | 1.125 |
| 1.100 | 5.500 | 1.350 | 1.929 | 1.600 | 1.333 | 1.850 | 1.088 |
| 1.150 | 3.833 | 1.400 | 1.750 | 1.650 | 1.269 | 1.900 | 1.056 |
| 1.200 | 3.000 | 1.450 | 1.611 | 1.700 | 1.214 | 1.950 | 1.026 |
| 1.250 | 2.500 | **1.500** | **1.500** | 1.750 | 1.167 | **2.000** | **1.000** |

Note: Given the form dimension $D_f$, the boundary dimension $D_b$ value can be yielded with equation (14).

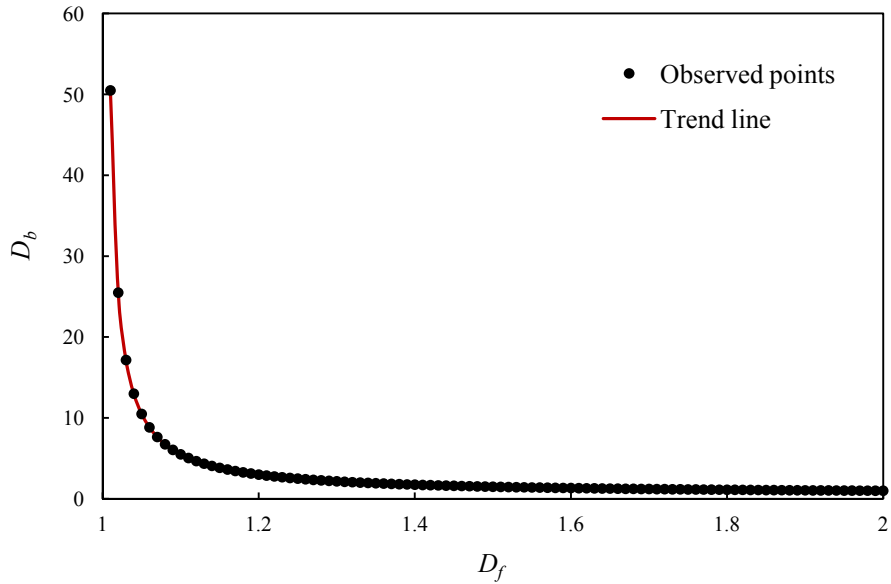

**Figure 1 The hyperbolic relation between the form dimension and boundary dimension**

[**Note:** The data points and line are created using equation (14). This curve suggests that if the form dimension $D_f$ value increase, the boundary dimension $D_b$ value will decrease accordingly. If the $D_f$ value is less than 4/3, the $D_b$ value will have no geometrical meaning.]

The boundary dimension differs from the form dimension where topology is concerned. The boundary dimension corresponds to the 1-dimensional space (the topological dimension is 1), while the form dimension corresponds to the 2-dimensional space (the topological dimension is 2). In other words, the boundary dimension is based on a fractal line resulting from a 1-dimensional initiator, while the form dimension is based on a fractal plane proceeding from a 2-dimensional initiator. Generally speaking, the form dimension should be greater than the boundary dimension, i.e., $D_f > D_b$. However, in theory, if and only if the form dimension is greater than 1.5, the $D_f$ value will be greater than the $D_b$ value. This seems to imply that the form dimension should come between 1.5 and 2 ($1.5 \leq D_f \leq 2$), which is consistent with proper scale of the radial dimension (Chen, 2013). Accordingly, the boundary dimension should come between 1 and 1.5 ($1 \leq D_b \leq 1.5$). If $D_l$ were the boundary dimension, the results in Table 5 might be inexplicable; to the contrary, if $D_b$ is regarded as the boundary dimension, the $D_f$ values are acceptable with reference to equation (14).

The form dimension values depend to a degree on the scopes of "viewfinding" (study area) on digital maps (Chen, 2012). If the study area is fixed far greater than the urban area within the exact/real urban boundary, the form dimension estimated will be low, or even lower than 1 (Shen, 2002). If we try to find an exact urban boundary, and define a proper study area by referring to its variable boundary, the fractal dimension values will be more reasonable (Benguigui *et al*, 2000). If the study area (viewfinding scope) matches the urban area, and the fractal dimension value is still very low, then the boundary line will be very complicated and zigzag, similar to the space-filling curves such as Hilbert's curve and Morton sequence of N-tree (Figure 2). The space-filling curves were employed to show the principle of recursive subdivision of geographical space (Goodchild and Mark, 1987). They are fractals with dimension equal to 2. The less an urban built-up area develops, the more an urban boundary line will fill. The hyperbolic relation between the boundary dimension and form dimension reminds us of the patterns of limited diffusion aggregation (DLA). The DLA models have been adopted to simulate urban growth and form (Batty *et al*, 1989; Chen, 2012; Fotheringham *et al*, 1989; Murcio and Rodríguez-Romo, 2009). In many cases, for DLA models, the lower form dimension suggests the higher boundary dimension.

The scaling relation between the linear scale and the number of nonempty boxes can be demonstrated to be a spatial correlation function. The fractal dimension of urban form is in fact a generalized correlation dimension. For a mature city, in the case of proper measurement, the form dimension value falls between 1.5 and 2 (Table 4). If the form dimension $D_f$=1.5, the corresponding spectral exponent will be $\beta$=1, which implies the $1/f$ noise (Chen, 2013). The indications of self-organized criticality comprise fractals, Zipf's law, and $1/f$ noise (Bak, 1996; Batty and Xie, 1999; Chen and Zhou, 2008). In this sense, the value of $D_f$=1.5 seems to suggest a self-organized critical state of urban evolution. If $D_f$<1.5, the spatial correlation function will be in inverse proportion to the distance $r$. This suggests the spatial centripetal force (the strength of concentration) has the advantage over the centrifugal force (the strength of deconcentration); therefore, urban development should fill in vacant space (spare land, vacant land) or even open space inwards. In this instance, city planning should be focused on internal space of a city (esp., city's proper). On the contrary, if $D_f$>1.5, the spatial correlation function will be in direct proportion to the distance $r$. This suggests the spatial centrifugal force has the advantage over the centripetal force; therefore, urban development should grow outwards, and outskirts are gradually occupied by structures, outbuildings, and service areas. In this case, city planning should be focused on external space of a city such as suburbs, or even exurbs (Chen, 2013).

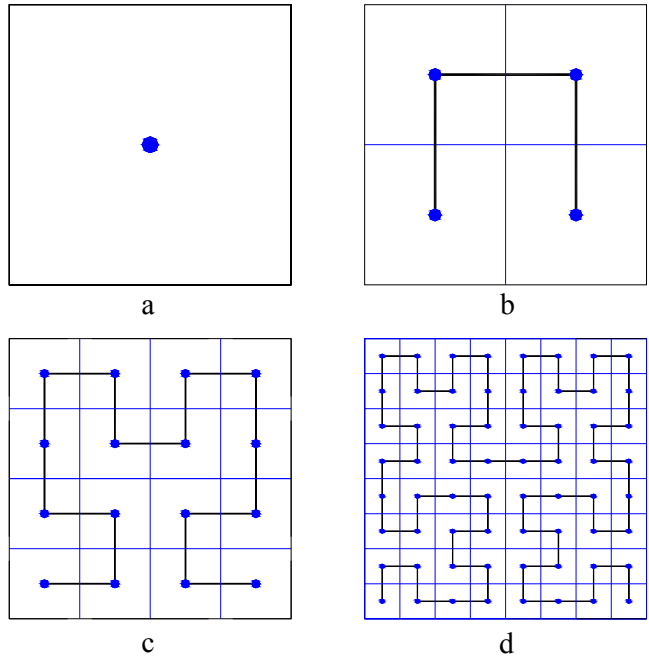

a. Hilbert's space-filling curve

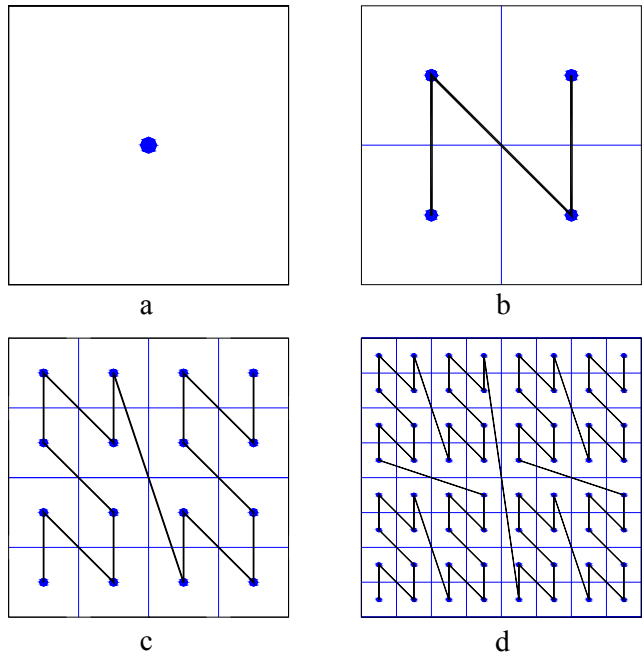

b. Morton sequence of N-tree

**Figure 2 Peano curve or space-filling tree**

A fractal is a scale-free phenomenon that bears no characteristic length in spatial measurement. However, the fractal dimension has its characteristic scale. There seems to be a best value for form dimension. According to the principles of fractal sets (Vicsek, 1989), we can define a fractal dimension of the intersection of urban field (fractal form) and envelope (fractal boundary) such as

$$D_i = D_b + D_f - d,  \tag{20}$$

where $D_i$ refers to the intersection dimension, and $d=2$ to the Euclidean dimension of the embedding space in which the structural fractal (urban area) and textural fractal (urban boundary) exist. The intersection of urban area and urban boundary is a fractal point set. If $D_f$=1.5, then $D_b$=1.5, and we will have $D_i$=1, which suggests degeneration of the fractal point set. On the other, If $D_f$=2, then $D_b$=1, also we will have $D_i$=1, this suggests degeneration of the city fractals, including fractal urban form and boundary line. If and only if $1<D_f<2$, we will have $D_i<1$ (Figure 3). This implies that there may be an optimum structure for the fractal point set, and the best fractal point set may suggest the optimized structure of fractal urban from. Substituting equations (15) into equations (20) yields

$$D_i = \frac{2D_f^2 - 5D_f + 4}{2(D_f - 1)}. \tag{21}$$

On the right side of equations (21), the numerator is a parabola. This suggests that there must be an extreme value for $D_i$. Taking derivative of $D_i$ with respect to $D_f$ gives

$$\frac{dD_i}{dD_f} = \frac{2D_f^2 - 4D_f + 1}{2(D_f - 1)^2}. \tag{22}$$

In terms of the condition of extreme value, we have a quadratic equation of the form dimension such as

$$2D_f^2 - 4D_f + 1 = 0. \tag{23}$$

This is a polynomial equation of the second degree. The roots can be given by the quadratic formula such as

$$D_f = \frac{4 \pm \sqrt{4^2 - 4*2*1}}{2*2} = 1 \pm \frac{\sqrt{2}}{2}. \tag{24}$$

The two roots are $D_f^*$=1+$2^{1/2}$/2≈1.7071 and $D_f^{**}$=1-$2^{1/2}$/2=2-$D_f^*$≈0.2929, respectively. The first root, $D_f^*$, is valid, while the second root, $D_f^{**}$, is not within the proper range of the form dimension value. Therefore, the best value of the form dimension is about $D_f$=1.7071, and the corresponding boundary dimension is around $D_b$=1.2071 in terms of equation (14). The minimum intersection dimension is $D_i$≈0.9142. This result lends further support to the suggestion of Batty and Longley (1994) that $D_f$=1.7 is a special value for the fractal dimension of urban form.

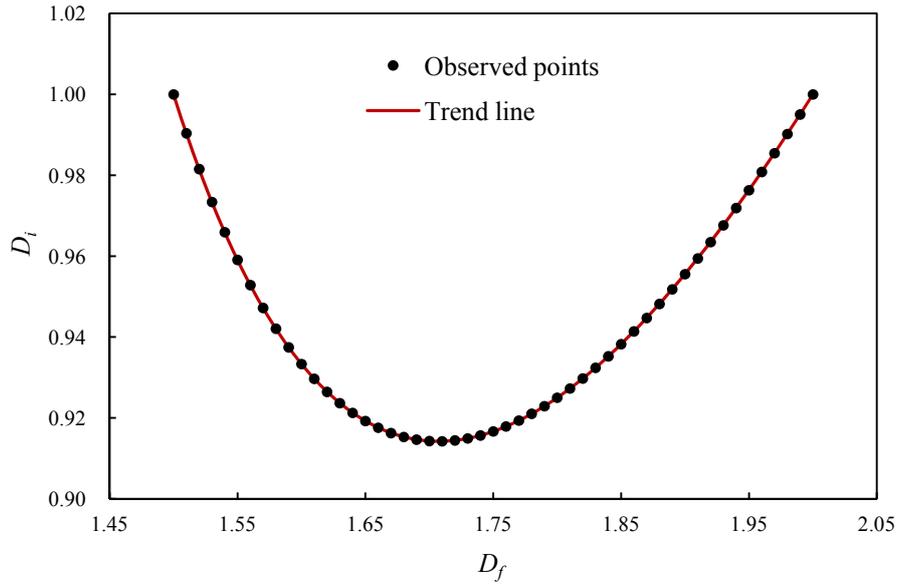

**Figure 3 The relation between the form dimension and the fractal parameter of the intersection of urban field and envelope**

[**Note:** The data points and line are generated using equation (21). This curve suggests that if the form dimension $D_f$ value increase, the fractal dimension of the intersection of urban field and envelope $D_i$ value will first decrease and then increase. If the $D_f$ value is close to 1.7071, the $D_i$ value will reach the minimum value.]

# 5 Conclusions

The fractal dimension study in this article is based on the box-counting method. Both the form dimension and boundary dimension are of box-counting dimension. Two scaling relations are employed. One is the mass (area/perimeter)-scale scaling, and the other, the area-perimeter scaling. The fractal dimension relations and fractal parameter estimation formulae are derived from the systems of scaling equations. A key lies in that a city fractal as a kind of prefractal is formally different from a mathematical fractal. For a city fractal, the area-perimeter scaling presents a power function indicating direct variations, while for a mathematical fractal, the area-perimeter scaling takes on a power law indicating inverse variations.

The mains of this paper can be summarized as follows.

First, and there exists a hyperbolic relation between the form dimension and boundary dimension of cities. The higher form dimension implies the lower boundary dimension and *vice versa*. By using the hyperbolic relation equation, we can estimate the form dimension if the boundary dimension is known, or we can estimate the boundary dimension if the form dimension is known. Both the form dimension and boundary dimension can be estimated through the area-perimeter scaling exponent. A set of practical fractal dimension formulae can be derived from the area-perimeter scaling and the box-counting method. The formulae are useful in future studies on fractal urban form and the related or similar studies on other physical or social systems.

Second, there are proper scale of values for the form dimension and boundary dimension. According to the hyperbolic relation, the boundary dimension comes between 1 and 3/2, and the form dimension comes between 3/2 and 2. If the form dimension is less than 4/3, the boundary dimension will be greater than 2 in theory, and this is abnormal. When the form dimension range from 4/3 to 3/2, the boundary dimension will vary from 3/2 to 2. If and only if the form dimension is greater than 3/2, the form dimension will be greater than the boundary dimension. The proper scale of the box dimension is consistent with that of the radial dimension of urban form.

Third, there are three points within the numerical interval of the form dimension, which ranges from 0 to 2. The first is $D_f=4/3$, this seems to be the lower limit of the form dimension, and the upper limit is $D_f=2$. The second is $D_f=3/2$, this seems to be a threshold value for urban growth. The form dimension $D_f=3/2$ implies a spectral exponent $\beta=1$, which in turn suggests a self-organized critical state of urban evolution. If the form dimension is less than 3/2, urban development should fill in vacant space or even open space inwards. On the other hand, if the form dimension is greater than 3/2, urban development should grow outwards, and outskirts are gradually occupied by structures, outbuildings, and service areas. The third is $D_f \approx 1.7071$, and this seems to be the best values of the form dimension indicating structural optimization of urban form.

**Acknowledgements**

This research was financially supported by the National Natural Science Foundation of China (Grant No. 41171129). I am grateful to two anonymous reviewers whose interesting comments were very helpful in improving the paper's quality.

# References

Addison PS (1997). *Fractals and Chaos: An Illustrated Course*. Bristol and Philadelphia: Institute of Physics Publishing

Ariza-Villaverde AB, Jimenez-Hornero FJ, De Rave EG (2013). Multifractal analysis of axial maps applied to the study of urban morphology. *Computers, Environment and Urban Systems*, 38: 1-10

Batty M, Fotheringham AS, Longley PA (1989). Urban growth and form: scaling, fractal geometry and diffusion-limited aggregation. *Environment and Planning A*, 21(11): 1447-1472

Batty M, Longley P (1987). Urban shapes as fractals. *Area*, 19(3): 215–221

Batty M, Longley PA (1988). The morphology of urban land use. *Environment and Planning B: Planning and Design*, 15(4): 461-488

Batty M, Longley PA (1994). *Fractal Cities: A Geometry of Form and Function*. London: Academic Press

Batty M, Xie Y (1999). Self-organized criticality and urban development. *Discrete Dynamics in Nature and Society*, 3(2-3): 109-124

Benguigui L, Blumenfeld-Lieberthal E, Czamanski D (2006). The dynamics of the Tel Aviv morphology. *Environment and Planning B: Planning and Design*, 33(2): 269-284

Benguigui L, Czamanski D, Marinov M, Portugali Y (2000). When and where is a city fractal? *Environment and Planning B: Planning and Design*, 27(4): 507–519

Chen YG (2010). Characterizing growth and form of fractal cities with allometric scaling exponents. *Discrete Dynamics in Nature and Society*, Vol. 2010, Article ID 194715, 22 pages

Chen YG (2011). Derivation of the functional relations between fractal dimension and shape indices of urban form. *Computers, Environment and Urban Systems*, 35(6): 442–451

Chen YG (2012). Fractal dimension evolution and spatial replacement dynamics of urban growth. *Chaos, Solitons & Fractals*, 45 (2): 115–124

Chen YG (2013). Fractal analytical approach of urban form based on spatial correlation function. *Chaos, Solitons & Fractals*, 49(1):47-60

Chen YG, Feng J (2012). Fractal-based exponential distribution of urban density and self-affine fractal forms of cities. *Chaos, Solitons & Fractals*, 45(11):1404–1416

Chen YG, Zhou YX (2006). Reinterpreting central place networks using ideas from fractals and self-organized criticality. *Environment and Planning B: Planning and Design*, 33(3): 345-364

Chen YG, Zhou YX (2008). Scaling laws and indications of self-organized criticality in urban systems. *Chaos, Solitons & Fractals*, 35(1): 85-98

Cheng Q (1995). The perimeter-area fractal model and its application in geology. *Mathematical Geology*, 27 (1): 69-82

De Keersmaecker M-L, Frankhauser P, Thomas I (2003). Using fractal dimensions for characterizing intra-urban diversity: the example of Brussels. *Geographical Analysis*, 35(4): 310-328

Feder J (1988). *Fractals*. New York: Plenum Press

Feng J, Chen YG (2010). Spatiotemporal evolution of urban form and land use structure in Hangzhou, China: evidence from fractals. *Environment and Planning B: Planning and Design*, 37(5): 838-856

Fotheringham S, Batty M, Longley P (1989). Diffusion-limited aggregation and the fractal nature of urban growth. *Papers of the Regional Science Association*, 67(1)**:** 55-69

Frankhauser P (1994). *La Fractalité des Structures Urbaines (The Fractal Aspects of Urban Structures)*. Paris: Economica

Frankhauser P (1998). The fractal approach: A new tool for the spatial Analysis of urban agglomerations. *Population: An English Selection*, 10(1): 205-240 [New Methodological Approaches in the Social Sciences]

Goodchild MF, Mark DM (1987). The fractal nature of geographical phenomena. *Annals of Association of American Geographers*, 77(2): 265-278

Imre AR (2006). Artificial fractal dimension obtained by using perimeter-area relationship on digitalized images. *Applied Mathematics and Computation*, 173 (1): 443-449

Imre AR, Bogaert J (2004). The fractal dimension as a measure of the quality of habitat. *Acta Biotheoretica*, 52(1): 41-56

Kaye BH (1989). *A Random Walk Through Fractal Dimensions*. New York: VCH Publishers

Lee Y (1989). An allometric analysis of the US urban system: 1960–80. *Environment and Planning A*, 21(4): 463–476

Longley PA, Batty M (1989a). On the fractal measurement of geographical boundaries. *Geographical Analysis*, 21 (1): 47-67

Longley PA, Batty M (1989b). Fractal measurement and line generalization. *Computer & Geosciences*,

15(2): 167-183

Longley PA, Batty M, Shepherd J (1991). The size, shape and dimension of urban settlements. *Transactions of the Institute of British Geographers (New Series)*, 16(1): 75-94

Lovejoy S, Schertzer D, Tsonis AA (1987). Functional box-counting and multiple elliptical dimensions in rain. *Science*, 235: 1036-1038

Lu Y, Tang J (2004). Fractal dimension of a transportation network and its relationship with urban growth: a study of the Dallas-Fort Worth area. *Environment and Planning B: Planning and Design*, 2004, 31(6): 895-911

Lung CW, Mu ZQ (1988). Fractal dimension measured with perimeter-area relation and toughness of materials. *Physical Review B*, 38 (16): 11781-11784

Mandelbrot BB (1977). *Fractals*: *Form*, *Chance*, *and Dimension*. New York: W. H. Freeman and Company

Mandelbrot BB (1982). *The Fractal Geometry of Nature.* New York: W. H. Freeman and Company

Murcio R, Rodríguez-Romo S (2009). Colored diffusion-limited aggregation for urban migration. *Physica A: Statistical Mechanics and its Applications*, 388(13): 2689–2698

Murcio R, Rodríguez-Romo S (2011). Modeling large Mexican urban metropolitan areas by a Vicsek Szalay approach. *Physica A: Statistical Mechanics and its Applications*, 390(16): 2895–2903

Shen G (2002). Fractal dimension and fractal growth of urbanized areas. *International Journal of Geographical Information Science*, 16(5): 419-437

Song Y, Wang SJ, Ye Q, Wang XW (2012). Urban spatial morphology characteristic and its spatial differentiation of mining city in China. *Areal Research and Development*, 31(1):45-39 [In Chinese]

Takayasu H (1990). *Fractals in the Physical Sciences*. Manchester: Manchester University Press

Terzi F, Kaya HS (2011). Dynamic spatial analysis of urban sprawl through fractal geometry: the case of Istanbul. *Environment and Planning B: Planning and Design*, 38(1): 175-190

Thomas I, Frankhauser P, Biernacki C (2008). The morphology of built-up landscapes in Wallonia (Belgium): A classification using fractal indices. *Landscape and Urban Planning*, 84(2): 99-115

Thomas I, Frankhauser P, De Keersmaecker M-L (2007). Fractal dimension versus density of built-up surfaces in the periphery of Brussels. *Papers in Regional Science*, 86(2): 287-308

Thomas I, Frankhauser P, Frenay B, Verleysen M (2010). Clustering patterns of urban built-up areas

with curves of fractal scaling behavior. *Environment and Planning B: Planning and Design*, 37(5): 942- 954

Vicsek T (1989). *Fractal Growth Phenomena*. Singapore: World Scientific Publishing Co.

Wang XS, Liu JY, Zhuang DF, Wang LM (2005). Spatial-temporal changes of urban spatial morphology in China. *Acta Geographica Sinica*, 60(3): 392-400 [In Chinese]

White R, Engelen G (1993). Cellular automata and fractal urban form: a cellular modeling approach to the evolution of urban land-use patterns. *Environment and Planning A*, 25 (8): 1175-1199

White R, Engelen G (1994). Urban systems dynamics and cellular automata: fractal structures between order and chaos. *Chaos, Solitons & Fractals*, 4(4): 563-583